\documentclass[cameraready]{Interspeech}

\title{An Empirical Analysis of Task-Induced Encoder Bias\\in Fr\'{e}chet Audio Distance}

\author[affiliation={}]{Wonwoo}{Jeong}

\address{Dept. of Computer Science and Engineering, Sogang University, South Korea}

\email{jeongwonwoo@sogang.ac.kr}


\usepackage{multirow}

\begin{document}

\maketitle

\begin{abstract}
Fr\'{e}chet Audio Distance (FAD) is the de facto standard for evaluating text-to-audio generation, yet its scores depend on the underlying encoder's embedding space. An encoder's training task dictates which acoustic features are preserved or discarded, causing FAD to inherit systematic task-induced biases. We decompose evaluation into Recall, Precision, and Alignment (split into semantic and structural dimensions), using log-scale normalization for fair cross-encoder comparison. Controlled experiments on six encoders across two datasets reveal a four-axis trade-off: reconstruction-based AudioMAE leads precision sensitivity; ASR-trained Whisper dominates structural detection but is blind to signal degradation; classification-trained VGGish maximizes semantic detection but penalizes legitimate intra-class variation. Since no single encoder is a universal evaluator, future metrics must shift toward evaluation-native encoders intrinsically aligned with human perception.
\end{abstract}

\section{Introduction}

Text-to-audio (TTA) generation has advanced rapidly with diffusion-based and language-model-based architectures~\cite{liu2023audioldm, kreuk2023audiogen, huang2023makeanaudio, copet2024musicgen, ghosal2023tango, liu2024audioldm2}, intensifying the need for reliable automatic evaluation. Fr\'{e}chet Audio Distance (FAD)~\cite{kilgour2019fad}, adapted from FID~\cite{heusel2017fid}, computes distributional distance between real and generated audio in a pretrained encoder's embedding space and has become the standard benchmark metric~\cite{copet2024musicgen, yang2023diffsound, agostinelli2023musiclm, evans2024stable}. However, FAD scores can diverge from human auditory judgments~\cite{gui2024adapting, vinay2022evaluating}---a limitation shared with its visual counterpart FID~\cite{jayasumana2024rethinkingfid} and one that undermines its reliability as a perceptual proxy.

While FAD's Gaussian assumption~\cite{chung2025kad} and sample size sensitivity~\cite{gui2024adapting} introduce significant limitations, the dominant and least-studied source of perceptual divergence remains the \emph{training task} of the encoder. The task explicitly determines feature preservation in the embedding space~\cite{parmar2022aliased}: an ASR encoder~\cite{radford2023whisper} abstracts away pitch and timbre; a classification encoder~\cite{hershey2017vggish} collapses temporal structure; and a codec encoder~\cite{defossez2022encodec} exhibits low sensitivity to inter-frame ordering. Consequently, distortions falling into an encoder's invariance set yield negligible FAD variations regardless of perceptual severity. Prior work has noted this encoder-dependent variability~\cite{gui2024adapting}, but it remains unclear exactly which features each encoder discards.

\begin{figure}[t]
  \centering
  \includegraphics[width=0.88\columnwidth]{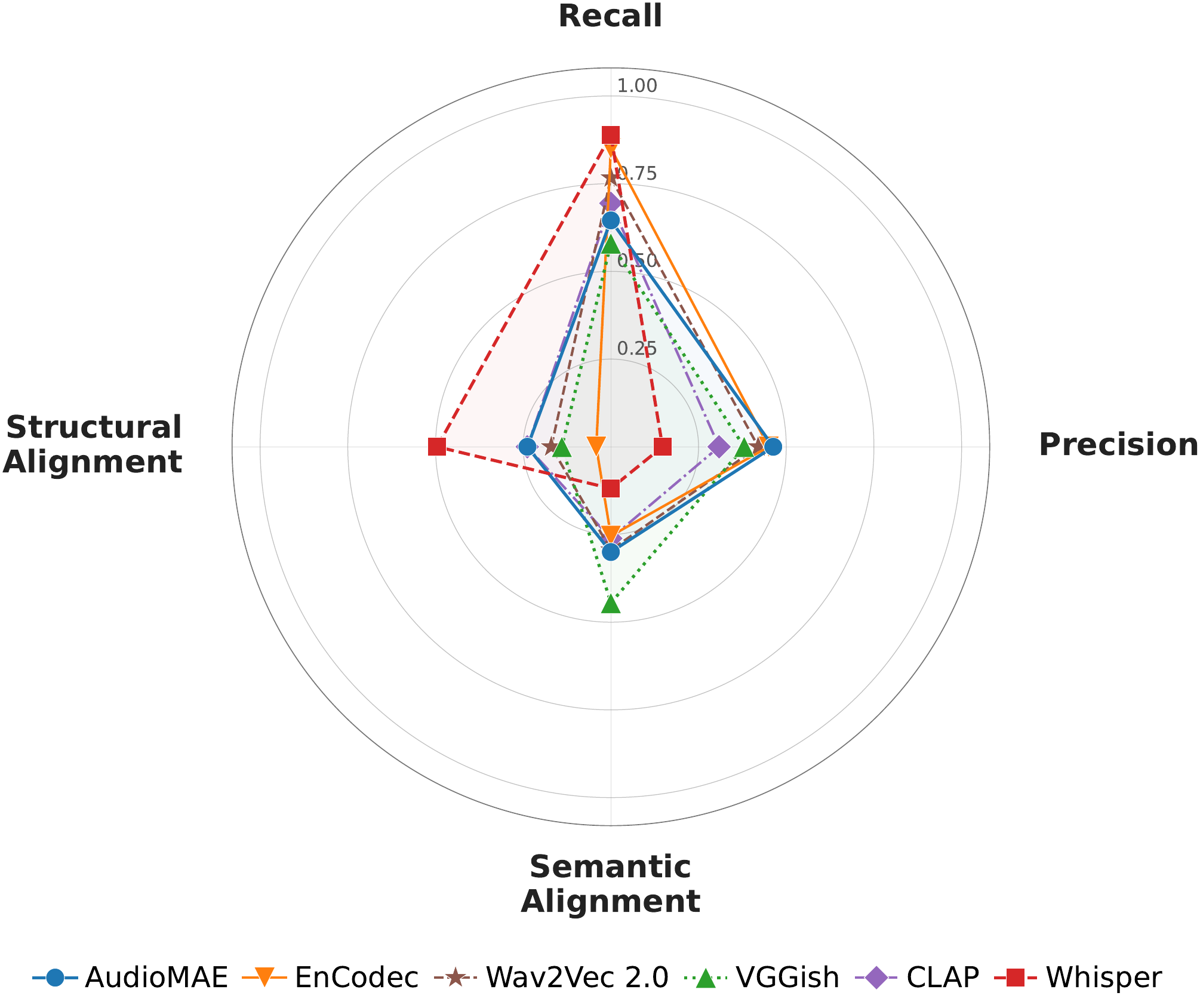}
  \caption{Four-axis trade-off (Outermost is better). AudioMAE leads Precision; Whisper captures Structural but is invariant to signal degradation; VGGish leads Semantic but limits Recall.}
  \label{fig:radar}
\end{figure}

To map these blind spots, we decompose evaluation into three axes: \textbf{Recall}, \textbf{Precision}, and \textbf{Alignment} (semantic, structural), and employ \textbf{log-scale self-reference normalization} as an analytical tool for fair cross-encoder comparison.

\newpage
Controlled experiments on six encoders and two datasets reveal a \emph{four-axis trade-off} (Figure~\ref{fig:radar}): AudioMAE achieves the highest precision sensitivity; Whisper dominates structural detection but exhibits marginal sensitivity to signal degradation; VGGish leads semantic alignment but disproportionately penalizes recall. Because every encoder's training task induces a distinct invariance set, no single tested encoder functions as a universal evaluator---a finding that underscores the need for evaluation-native encoders whose embedding spaces are intrinsically aligned with human perception.

\section{Analytical Methodology}

\subsection{Fr\'{e}chet Audio Distance and Encoder Projection}

Let $\mathcal{X}$ be the input audio space and $\Phi : \mathcal{X} \to \mathcal{Z} \subset \mathbb{R}^d$ be a pretrained encoder, where $\mathcal{Z}$ denotes the $d$-dimensional embedding space induced by the training task. FAD measures the 2-Wasserstein distance between reference and generated embedding distributions in $\mathcal{Z}$, modeled as Gaussians $\mathcal{N}(\mu_r, \Sigma_r)$ and $\mathcal{N}(\mu_g, \Sigma_g)$ estimated from the reference set $\mathcal{R}$ and generated set $\mathcal{G}$, respectively:
\begin{equation}
\mathrm{FAD} = {\Vert\mu_r {-} \mu_g\Vert}^2 + \mathrm{tr}\big(\Sigma_r {+} \Sigma_g {-} 2(\Sigma_r\Sigma_g)^{1/2}\big)
\label{eq:fad}
\end{equation}
While FAD constitutes a valid distributional metric within $\mathcal{Z}$, its capacity to represent perceptual distance in $\mathcal{X}$ is fundamentally constrained by the structure of $\Phi$. To analyze this limitation, let $\mathcal{G}_{\mathrm{pert}}$ be a perturbed set where each reference sample $x \in \mathcal{R}$ is replaced by its transformed counterpart $\tilde{x} \in \mathcal{X}$ subjected to a specific perceptual perturbation (e.g., temporal shuffling, additive noise). If the encoder is invariant to this perturbation such that $\Phi(\tilde{x}) \approx \Phi(x)$ for all $x \in \mathcal{R}$, the embedding distributions in $\mathcal{Z}$ become statistically indistinguishable, rendering FAD insensitive to the transformation. With a slight abuse of notation, we refer to the set of such perturbations as the encoder's \emph{approximate invariance set}:
\begin{equation}
\Phi(\tilde{x}) \approx \Phi(x) \enspace\Longrightarrow\enspace \mathrm{FAD}(\mathcal{R}, \mathcal{G}_{\mathrm{pert}}) \to 0
\label{eq:kernel}
\end{equation}
Consequently, FAD does not measure absolute perceptual distance, but rather the divergence \emph{projected exclusively onto the specific subspace} preserved by the encoder's training task. This invariance set directly dictates the range of variation an encoder \emph{accommodates}---thereby defining its inherent evaluation biases. Identifying its composition for each encoder is the central goal of our analysis.

\subsection{The Recall--Precision--Alignment Analysis Axes}

A conditional generative model should satisfy three desiderata: covering the full support of the target distribution (\emph{Recall}), confining samples strictly within the support of the target distribution (\emph{Precision}), and localizing these samples to the specific conditional distribution dictated by the prompt (\emph{Alignment}). Sajjadi et al.~\cite{sajjadi2018precision} demonstrated that single distributional metrics conflate these dimensions, proposing a Precision--Recall decomposition later refined by kNN~\cite{kynkaanniemi2019improved} and density/coverage analyses~\cite{naeem2020density}. We adopt this decomposition \emph{as an analytical lens}---not the kNN-based computation---to systematically diagnose which perceptual dimensions each encoder's FAD captures or penalizes.

Moreover, TTA is a \emph{conditional} generation task: the target distribution in~$\mathcal{Z}$ is not a monolithic mode but rather a collection of condition-specific clusters. If a model produces audio that misaligns with its prompt, the generated embeddings land in a disjoint region of~$\mathcal{Z}$, displacing~$\mu_g$ from~$\mu_r$ and thereby inflating the mean-distance term in Eq.~\ref{eq:fad}. Thus, Alignment constitutes a third axis inherently captured by FAD, provided the encoder~$\Phi$ preserves the relevant conditional structure. Specifically, because a prompt dictates both the semantic identity of a sound source and the temporal ordering of events, an encoder that captures one dimension might still be completely blind to the other. To prevent a single alignment metric from obscuring these distinct blind spots, we further split Alignment into semantic and structural components, yielding a four-axis evaluation profile:

\textbf{Recall} measures tolerance to intra-class variation---the degree of stylistic variability (e.g., mild pitch or tempo differences) an encoder permits before registering a distributional shift. We probe whether an encoder actively \emph{penalizes} legitimate variation when FAD serves as an optimization target~\cite{friedman2023vendi}.

\textbf{Precision} quantifies sensitivity to perceptual artifacts---noise, bandwidth loss, reverberation coloring---that degrade sample fidelity regardless of the intended content.

\textbf{Alignment} measures consistency with the conditioning input. We subdivide this into \emph{Semantic Alignment} (preservation of spectral identity---source timbre, sound character) and \emph{Structural Alignment} (temporal ordering---chronological flow, event sequencing).

\subsection{Cross-Encoder Normalization for Comparative Analysis}

To enable fair comparison across encoders whose dynamic ranges span multiple orders of magnitude, we employ a log-scale self-reference normalization. For encoder~$e$ and perturbation condition~$\tau$ (e.g., ``white noise at SNR\,20\,dB''), let $\mathrm{FAD}^{(e)}(\tau) \equiv \mathrm{FAD}(\mathcal{R},\,\mathcal{G}_{\tau})$ denote the FAD computed via encoder~$e$ between the clean reference set and the set perturbed by~$\tau$:
\begin{equation}
S_{\mathrm{norm}}^{(e)}(\tau) = \frac{\log(1 + \mathrm{FAD}^{(e)}(\tau))}{\log(1 + \mathrm{FAD}_{\max}^{(e)})}
\label{eq:snorm}
\end{equation}
where $\mathrm{FAD}_{\max}^{(e)}$ is the maximum FAD observed for encoder~$e$ across our fixed perturbation suite, rendering the normalization deterministic. Since every encoder is exposed to identical perturbations, $\mathrm{FAD}_{\max}^{(e)}$ reflects each encoder's intrinsic dynamic range. The $\log(1{+}\cdot)$ transform provides a monotonic compressive mapping; our goal is comparative profiling of task-induced sensitivities, not absolute cross-encoder ranking. Using the 95th percentile instead of the maximum yields identical qualitative conclusions.

Figure~\ref{fig:fidelity} illustrates the necessity of this normalization. On a linear scale~(a), the broad dynamic range of EnCodec (reaching $148.8$ at SNR~$-5$\,dB) compresses the response trajectories of less sensitive encoders (e.g., VGGish, Whisper, and CLAP) into a visually indistinguishable baseline. After log-scale normalization~(b), the variations among all six encoders become distinctly observable. This transformation is justified on two grounds:

\begin{figure}[t]
  \centering
  \includegraphics[width=\columnwidth]{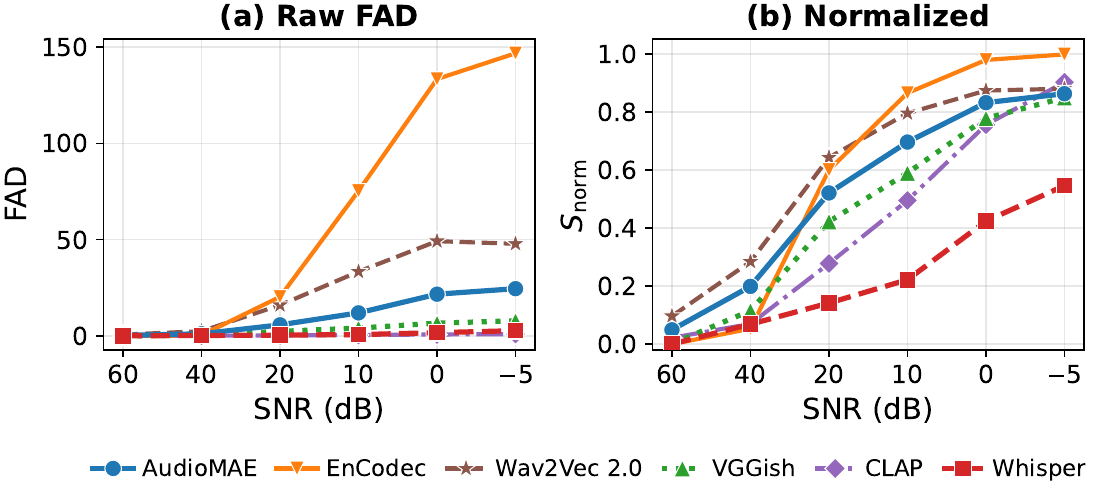}
  \caption{Precision response to white noise. (a)~Raw FAD (linear scale): dynamic-range disparity compresses low-sensitivity encoders into a visually indistinguishable baseline---this ``visual squashing'' motivates our normalization. (b)~After log-scale normalization ($S_{\mathrm{norm}}$), all six trajectories separate, revealing each encoder's distinct precision profile.}
  \label{fig:fidelity}
\end{figure}

\textbf{Mitigation of scale-disparity compression.}
FAD is unbounded; as Figure~\ref{fig:fidelity}(a) shows, dynamic ranges differ by over two orders of magnitude (EnCodec exceeds $148$ while CLAP barely reaches $1.0$). Linear normalization bounded by these extremes would compress the low-distortion regime ($\mathrm{FAD} {<} 10$) into a negligible interval ($S_{\mathrm{norm}} {<} 0.07$), rendering key encoder differences visually imperceptible.

\textbf{Discriminative resolution in the low-distortion regime.}
Standard linear scaling implicitly assumes constant perceptual sensitivity ($\partial\,\mathrm{Perception}\,/\,\partial\,\mathrm{Distance} \approx c$). In practice, the discriminative utility of the metric decays as distributional distance grows: the distinction between \emph{perfect} and \emph{slightly flawed} is critical, whereas the distinction between \emph{destroyed} and \emph{more destroyed} is negligible. The $\log(1{+}\cdot)$ transformation concentrates discriminative resolution in the low-distortion regime---a sensitivity gradient consistent with the Weber--Fechner law of human perception~\cite{weber1834, fechner1860}.

Each axis score averages $S_{\mathrm{norm}}$ over its assigned perturbations across both datasets, ensuring the reported profiles reflect domain-general encoder behavior rather than dataset-specific artifacts. We report $1 - \bar{S}_{\mathrm{norm}}$ for Recall to ensure directional consistency across the four-axis profile, while Figures~\ref{fig:fidelity}--\ref{fig:diversity} plot the raw sensitivity $S_{\mathrm{norm}}$ to detail the encoder's response trajectory.

\section{Experimental Setup}

\subsection{Datasets and Preprocessing}

We utilize \textbf{LibriSpeech test-clean}~\cite{panayotov2015librispeech} (2{,}620 utterances, variable length) and \textbf{ESC-50}~\cite{piczak2015esc} (2{,}000 environmental sounds, 5\,s each) to span the speech and general audio domains. Clips retain their original duration without padding or truncation: because FAD aggregates per-clip embeddings into set-level Gaussian statistics, duration variability is inherently neutralized between paired reference and perturbed sets. All audio is loudness-normalized to $-23$~LUFS (ITU-R~BS.1770-4) and resampled to each encoder's native rate.

\subsection{Encoders}

We evaluate six encoders spanning five training paradigms (Table~\ref{tab:encoders}), carefully chosen to probe complementary regions of the invariance--sensitivity landscape:

\textbf{Semantic encoders.}
Whisper~\cite{radford2023whisper} (ASR) represents linguistic structure preservation and provides an empirical ceiling for temporal sensitivity. CLAP~\cite{wu2023clap} maps audio into a shared text--audio space to probe cross-modal alignment transfer. VGGish~\cite{hershey2017vggish} provides a baseline for spectral template sensitivity.

\textbf{Acoustic encoders.}
AudioMAE~\cite{huang2022audiomae} (masked reconstruction) establishes the empirical ceiling for signal-level precision sensitivity. EnCodec~\cite{defossez2022encodec} (neural codec) tests whether compression-oriented training introduces systematic frequency-band biases. Wav2Vec~2.0~\cite{baevski2020wav2vec} (self-supervised, related to HuBERT~\cite{hsu2021hubert} and BEATs~\cite{chen2022beats}) serves as a task-agnostic baseline. For transformer-based encoders (Whisper, AudioMAE, Wav2Vec~2.0), we extract the final hidden state; for EnCodec, we utilize the continuous encoder output prior to the residual vector quantizer. FAD requires a single clip-level embedding per sample. Following standard practice~\cite{kilgour2019fad}, we apply temporal mean-pooling to each frame-level encoder's output; CLAP natively produces a clip-level embedding via internal attention pooling. Holding the aggregation constant across all five frame-level encoders ensures that the observed sensitivity differences strictly isolate the training-task effect rather than the pooling mechanism (see Section~4.4). FAD is then computed between embeddings of the clean reference and perturbed sets via Gaussian statistics (Eq.~\ref{eq:fad}).

\setlength{\tabcolsep}{3pt}

\begin{table}[t]
  \caption{Audio encoders and their training tasks. Task type determines which acoustic features are preserved in~$\mathcal{Z}$.}
  \label{tab:encoders}
  \centering
  \small
  \begin{tabular}{llcc}
    \toprule
    \textbf{Encoder} & \textbf{Training Task} & \textbf{SR} & \textbf{Dim.} \\
    \midrule
    AudioMAE    & Masked Reconstruction            & 16k & 768 \\
    EnCodec     & Neural Audio Compression         & 24k & 128 \\
    Wav2Vec 2.0 & Contrastive Learning (SSL)       & 16k & 768 \\
    VGGish      & Audio Classification             & 16k & 128 \\
    CLAP        & Cross-modal Contrastive Learning & 48k & 512 \\
    Whisper     & Automatic Speech Recognition     & 16k & 1280 \\
    \bottomrule
  \end{tabular}
\end{table}

\setlength{\tabcolsep}{9pt}

\subsection{Perturbation Design}

To systematically probe the encoders' invariance sets, we design a suite of targeted perturbations, each mapped to a specific evaluation axis:

\textbf{Recall.} Mild pitch shift ($\pm 1, \pm 2$~st) and time stretch ($0.9{\times}, 1.1{\times}$) represent intra-class stylistic variations that should not trigger significant distributional shifts.

\textbf{Precision.} Signal-level degradations include additive white noise at SNR $\in \{60, 40, 20, 10, 0, {-}5\}$~dB, low-pass biquad filtering with cutoffs at $\{8000, 6000, 4000, 2000, 1000\}$~Hz, and reverberation with RT$_{60} \in \{0.1, 0.2, 0.25, 0.4, 0.5, 0.6, 0.8, 1.0, 2.0\}$~s.

\textbf{Semantic Alignment.} Severe distortions include pitch shift at $\pm 4, \pm 8$~st and spectral envelope manipulation (``formant shift'') at $1.3{\times}, 1.4{\times}$ with $F_0$ preserved. While originating in speech processing, this reshaping modifies the perceived physical dimensions and resonance of general audio sources, effectively inducing a categorical shift in source identity.

\textbf{Structural Alignment.} Macroscopic temporal disruptions are evaluated via time reversal and chunk shuffling at durations of $\{1000, 500, 250, 100\}$~ms. We apply \SI{10}{\milli\second} cross-fades to eliminate click artifacts and isolate structural effects.

All transformations are executed via standard DSP libraries. The demarcation between Recall and Semantic Alignment is operationally defined by general perceptual tendencies: mild $\pm$1--2\,st shifts typically represent natural expressive variation, whereas larger $\pm$4--8\,st shifts tend to alter the perceived source identity.

\section{Results and Discussion}

\subsection{The Four-Axis Trade-off}

As detailed in Table~\ref{tab:fad_scores} and Figure~\ref{fig:radar}, AudioMAE demonstrates the highest Precision, closely followed by EnCodec. Whisper exhibits an orthogonal profile, maximizing Structural Alignment and Recall while minimizing Precision. Conversely, VGGish maximizes Semantic Alignment at the expense of Recall. CLAP remains balanced but achieves no peak sensitivities. Two distinct mechanisms underlie this trade-off: the Semantic--Recall opposition is fundamental, as classification training collapses features onto tight class-mean clusters~\cite{papyan2020neural}, intrinsically penalizing legitimate intra-class variation; the Precision--Structural divergence is paradigm-specific, driven by the differing tasks of ASR versus reconstruction training rather than an intrinsic constraint.

\subsection{Recall: Asymmetry and Inflexible Sensitivity}

\begin{figure}[t]
  \centering
  \includegraphics[width=0.92\columnwidth]{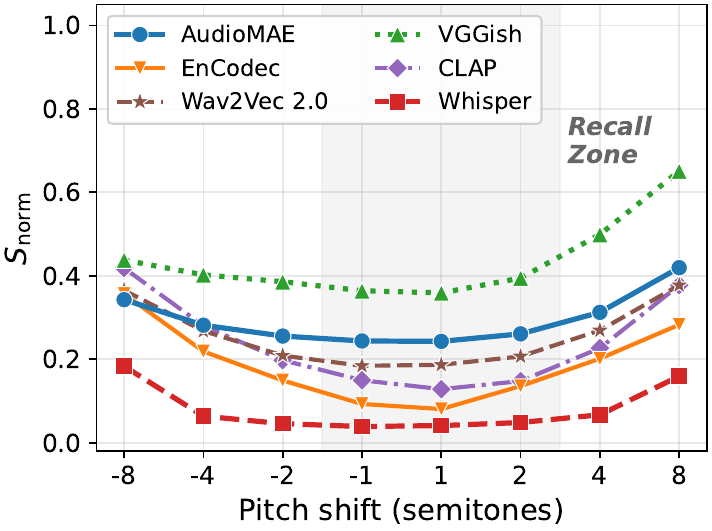}
  \caption{Pitch-shift trajectory ($-8$ to $+8$\,st). Shaded: recall zone ($\pm$1--2\,st). VGGish shows inflexible sensitivity at mild shifts; Whisper maintains the lowest recall-zone response.}
  \label{fig:diversity}
\end{figure}

Figure~\ref{fig:diversity} maps the full pitch-shift trajectory from $-8$ to $+8$\,st, revealing two salient phenomena. First, \emph{directional asymmetry}: VGGish exhibits a $0.6{\times}$ weaker response to downward than upward shifts, suggesting classification boundaries skewed toward higher frequencies. EnCodec shows a moderate downward bias ($1.6{\times}$), while AudioMAE, Wav2Vec~2.0, and Whisper remain approximately symmetric.

Second, \emph{inflexible sensitivity}: in the pitch recall zone ($\pm$1--2\,st, shaded), VGGish registers $S_{\mathrm{norm}} {=} 0.36$ at $+1$\,st---$9{\times}$ Whisper's $0.04$---because classification training interprets even minor spectral displacements as categorical deviations: \textbf{a \emph{recall trap}}. The contrast between near-uniform responses within $\pm$2\,st and divergent, asymmetric profiles at $\pm$4--8\,st empirically validates the perceptual-magnitude demarcation between Recall and Semantic Alignment (Section~3.3). This rigid sensitivity extends to the temporal domain: under mild time stretch ($0.9{\times}$--$1.1{\times}$), all encoders show disproportionately higher sensitivity than under comparable pitch shifts. Consequently, optimizing for single-encoder FAD risks constraining generative models to narrow spectral and temporal templates, actively penalizing natural variations that listeners typically accept~\cite{kumar2019melgan, kong2020diffwave}.

\begin{table}[t]
  \caption{Normalized R/P/A scores (Eq.~\ref{eq:snorm}). Recall: higher is more tolerant. Precision and Alignment: higher is more sensitive. \textbf{Bold}: best per column.}
  \label{tab:fad_scores}
  \centering
  \small
  \begin{tabular}{lcccc}
    \toprule
    \textbf{Encoder} & \textbf{Rec.} & \textbf{Prec.} & \textbf{Sem.} & \textbf{Struct.} \\
    \midrule
    AudioMAE     & 0.645          & \textbf{0.463} & 0.300 & 0.238 \\
    EnCodec      & 0.851          & 0.450          & 0.254 & 0.042 \\
    Wav2Vec 2.0  & 0.767          & 0.420          & 0.294 & 0.170 \\
    VGGish       & 0.580          & 0.380          & \textbf{0.445} & 0.140 \\
    CLAP         & 0.694          & 0.309          & 0.261 & 0.238 \\
    Whisper      & \textbf{0.889} & 0.147          & 0.119 & \textbf{0.495} \\
    \bottomrule
  \end{tabular}
\end{table}

\subsection{Precision: Threshold Behavior and Codec Blind Spots}

Whisper's suppressed precision sensitivity ($\bar{S}_{\mathrm{norm}} {=} 0.23$ for noise, $0.14$ for reverberation) is consistent with ASR training that incentivizes noise-robust representations to maintain transcription accuracy. In contrast, VGGish ($0.46$ noise, $0.44$ reverberation) and AudioMAE saturate rapidly, reflecting their acute sensitivity to signal-level degradation. EnCodec's low-pass response on LibriSpeech reveals a distinct anomaly: FAD jumps $32{\times}$ between 6\,kHz and 8\,kHz, as RVQ capacity concentrates sub-8\,kHz~\cite{defossez2022encodec}---this discontinuity does not manifest on ESC-50, indicating a speech-specific bandwidth bias.

\subsection{Alignment: Content vs.\ Order}

\begin{figure}[t]
  \centering
  \includegraphics[width=\columnwidth]{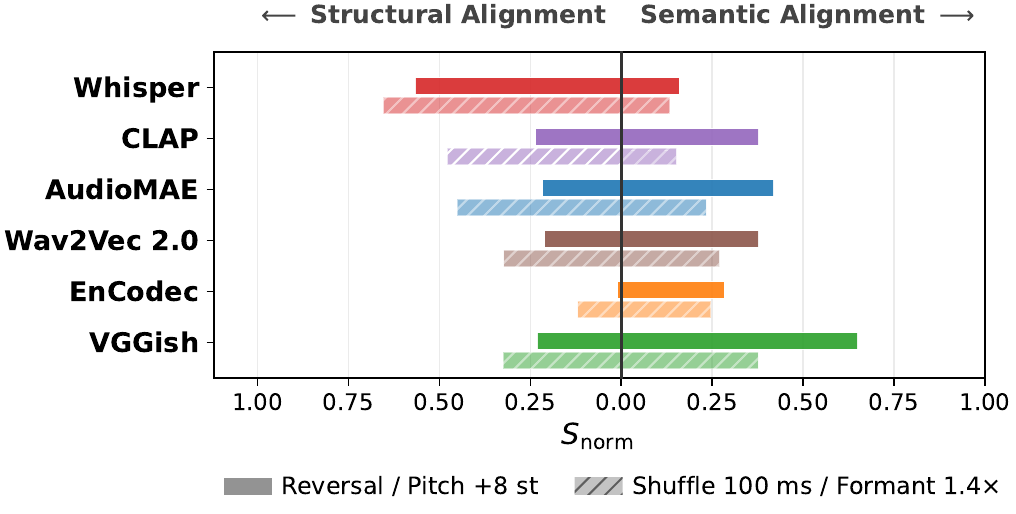}
  \caption{Diverging bar chart of Structural (left) vs.\ Semantic (right) sensitivity. Solid bars: Reversal / Pitch\,+8\,st; hatched bars: Shuffle\,100\,ms / Formant\,1.4$\times$. Whisper extends far left (structural-dominant); VGGish extends far right (semantic-dominant)---visually capturing the anti-correlation ($r{=}{-}0.67$).}
  \label{fig:blindspots}
\end{figure}

A Pearson correlation analysis between the Structural and Semantic scores (Table~\ref{tab:fad_scores}) across all six encoders reveals a strong anti-correlation ($r{=}{-}0.67$). Figure~\ref{fig:blindspots} illustrates this inverse relationship. Under identical mean-pooling, Whisper demonstrates pronounced sensitivity to structural disruptions while remaining largely invariant to semantic shifts; VGGish exhibits the inverse profile, whereas AudioMAE and EnCodec occupy intermediate positions. Because temporal aggregation is held constant across the five frame-level encoders, the $1.5{\times}$ sensitivity gap between Whisper and AudioMAE at the 100\,ms shuffle condition strictly isolates the training-task effect. Specifically, ASR training encodes sequential dependencies that persist through temporal pooling, whereas codec~\cite{defossez2022encodec} tasks yield order-invariant embeddings \emph{prior} to aggregation. This divergence demonstrates that no single evaluated encoder simultaneously captures both structural and semantic alignment.

\subsection{Implications and Limitations}

The four-axis trade-off yields direct practical consequences. Our analysis contextualizes previously reported human--metric discrepancies~\cite{gui2024adapting, vinay2022evaluating}: VGGish disproportionately penalizes natural variations that listeners accept, whereas reconstruction-based encoders largely fail to detect structural violations. Although multi-encoder aggregation might appear as a potential solution, the principled fusion of heterogeneous embedding spaces---differing in sample rate, dimensionality, and dynamic range---constitutes an open problem. Practitioners should therefore transition away from monolithic FAD reporting, explicitly disclosing the chosen encoder and selecting one aligned with their specific evaluation goals.

More fundamentally, relying on task-specific encoders limits FAD by design; their invariance sets are immutable artifacts of their training tasks. Unlike these models, human auditory perception integrates semantic identity and structural flow without mutually exclusive trade-offs. Overcoming this limitation requires shifting toward representations intrinsically aligned with human perception. Rather than serving as a direct optimization target, the presented R/P/A decomposition functions as an analytical lens to diagnose these blind spots and audit future metrics for balanced sensitivities.

While our approach effectively isolates task-induced biases, several limitations remain. First, mapping the analytical R/P/A axes to human auditory judgments necessitates large-scale subjective testing (e.g., FAD--MOS correlation). Second, to avoid over-generalization, the observed trends should be verified across a broader range of architectures within each paradigm (e.g., HuBERT~\cite{hsu2021hubert}, AST~\cite{gong2021ast}). Furthermore, while our DSP-based perturbations isolate specific features, real-world generative artifacts are highly entangled, substantially complicating diagnostic separation. Finally, extending this analysis to the music domain---where harmony, rhythm, and timbre interact intricately---is critical for comprehensive metric evaluation.

\section{Conclusion}

FAD is inherently a \emph{task-shaped projection}, capturing only the divergence its encoder's training task preserves. Our R/P/A decomposition reveals distinct invariance profiles---AudioMAE leads precision but not structure; Whisper captures temporal order yet is blind to signal degradation; VGGish maximizes semantic sensitivity while penalizing intra-class variation.

Since these blind spots are immutable training artifacts, no single encoder can serve as a universal evaluator, requiring researchers to explicitly disclose their encoder selection. Reconciling FAD with human auditory judgment necessitates embedding spaces whose geometric structure is aligned with perceptual sensitivity rather than dictated by a single training task---a direction for which our R/P/A decomposition provides an analytical lens.

\newpage
\section{Acknowledgments}

\ifcameraready
The author acknowledges the computational resources provided by Sogang University.
\else
The authors
\fi

\section{Generative AI Use Disclosure}

Generative AI tools were used only for language editing and polishing; they were not used to produce a significant portion of this manuscript.


\end{document}